\documentclass[letter-sized,journal]{IEEEtran}
\usepackage{amsmath,amsfonts}
\usepackage{algorithmic}
\usepackage{algorithm}
\usepackage{array}
\usepackage[caption=false,font=normalsize,labelfont=sf,textfont=sf]{subfig}
\usepackage{textcomp}
\usepackage{stfloats}
\usepackage{url}
\usepackage{verbatim}
\usepackage{graphicx}
\usepackage{cite}
\usepackage{nicefrac}
\usepackage{xcolor, soul}
\usepackage{svg}
\sethlcolor{yellow}
\usepackage{color}

\usepackage{tikz}
\usepackage{pgfplots}

\begin{document}

\title{Transformer Masked Autoencoders for Next-Generation Wireless 
		Communications: Architecture and Opportunities
	\author{Abdullah Zayat, Mahmoud A. Hasabelnaby,~\textit{Graduate Student Member,~IEEE,} Mohanad Obeed,~\textit{Member,~IEEE,} Anas~Chaaban,~\textit{Senior Member,~IEEE}}	
	\thanks{A. Zayat, M. A. Hasabelnaby, and A. Chaaban are with the School of Engineering, University of British Columbia, Kelowna, BC V1V1V7, Canada.}
		\thanks{M. Obeed is with the Systems and Computer Engineering Department, Carleton University, Ottawa, ON K1S 5B6, Canada.}
}
\maketitle

\maketitle

\begin{abstract}
Next-generation communication networks are expected to exploit recent advances in data science and cutting-edge communications technologies to improve the utilization of the available communications resources. 
In this article, we introduce an emerging deep learning (DL) architecture, the transformer-masked autoencoder (TMAE), and discuss its potential in next-generation wireless networks. {We discuss the limitations} of current DL techniques in meeting the requirements of 5G and beyond 5G networks, and how the TMAE differs from the classical DL techniques {can potentially} address several wireless communication problems. {We highlight} various {areas} in next-generation mobile networks {which can be addressed using a TMAE}, including source and channel coding, estimation, and security. Furthermore, we demonstrate a case study showing how a TMAE can {improve} data compression {performance and complexity} compared to existing schemes. Finally, we discuss key challenges and open future research directions for deploying the TMAE in intelligent next-generation mobile networks.

\end{abstract}

\begin{IEEEkeywords}
6G, 5G, convolutional neural networks, deep learning, wireless communication, recurrent neural networks, transformer, masked autoencoder.
\end{IEEEkeywords}

\section{Introduction}

Next-generation (NG) mobile networks are increasingly calling for intelligent architectures that support massive connectivity, ultra-low latency, ultra-high reliability, high-quality of experience, high spectral and energy efficiency, and lower deployment costs \cite{9355403}. One way to meet these stringent requirements is to rethink traditional communication techniques by exploiting recent advances in artificial intelligence. 




Traditionally, functions such as waveform design, channel estimation, interference mitigation, and error detection and correction are developed based on {theoretical} models and assumptions. This traditional approach is not capable of adapting to new challenges introduced by emerging technologies. For instance, the pilot-based channel estimation technique, while efficient for MIMO systems with a few antennas and low-mobility users, is not efficient for massive multiple-input multiple-output (MIMO) systems or high-mobility users. 
Additionally, the plethora of communication protocols, technologies, and services that have been introduced to support the growing demand and diversity of use cases make it increasingly difficult to mathematically model {wireless networks.}
As such, optimizing {communication schemes} using mathematical models becomes extremely challenging and computationally complex, particularly with the integration of demanding 6G applications and services, such as the Metaverse, ubiquitous Extended Reality (XR), intelligent connected robotics, large-scale intelligent reconfigurable surfaces (IRSs), and ultra-massive MIMO (UM-MIMO) networks.

Due to this, the use of deep learning {(DL)} approaches has been proposed to solve wireless communications challenges due to their ability to adapt to dynamic environments, approximate complex models, and utilize data to improve  performance \cite{sun2019application}. Transformer{-enabled DL}, initially proposed for natural language processing (NLP) tasks \cite{10.5555/3295222.3295349}, opens the door for further advances in this area. The main advantage of transformers is their superior ability to learn complex dependencies between input features compared to classical deep neural networks (DNN) \cite{9319626}. The goal of this paper is to discuss a transformer-based architecture, the transformer masked autoencoder (TMAE), and its application in communications. Particularly, we start by highlighting some limitations of existing DNNs (Sec. II), then we explain the general transformer and the MAE architectures (Sec. III). Next, we present a use case where a TMAE enhances data compression by exploiting semantics
, which can significantly improve achievable rates in wireless networks (Sec. IV). Finally, we discuss opportunities for improving wireless communications using the TMAE (Sec. V) and summarize the takeaway message of the paper (Sec.~VI).  


\section{Classical DNN Limitations in NG Networks}

Recent advances in DL opened the possibility for designing intelligent mobile networks that learn to operate optimally using massive amounts of data \cite{sun2019application}, thus overcoming mathematical modeling and computational complexity challenges \cite{8663966}. 
Although DL offers advantages over mathematical models, it is not free of limitations. 
Some {common DNNs are discussed next, followed by their limitations in NG networks.} 


\subsection{Common DNN Architectures} 
{The most commonly used DNN architectures in wireless communications include the following:} 
\begin{itemize}
\item Multi-Layer Perceptions (MLP): An MLP is a feed-forward neural network (NN) that consists of at least three layers of fully-connected nodes: an input layer, a hidden layer, and an output layer. 
MLPs have been proposed to tackle various mobile network problems, such as beamforming and channel estimation \cite{8663966}. 
\item Convolutional Neural Networks (CNN): CNNs replace fully-connected layers with locally connected kernels that capture local correlations in data. This reduces the number of model parameters which simplifies training and reduces the risk of overfitting. As a result, CNNs often outperform MLPs in many applications such as computer vision (CV). CNNs have also been used in many wireless communications applications such as semantic communications \cite{sun2019application}, beamforming, channel estimation, and channel state information (CSI) feedback \cite{8663966}. 
\item Recurrent Neural Networks (RNN): The main difference between a typical NN (MLP or CNN) and an RNN is that an RNN has feedback connections in addition to feed-forward connections. These connections give RNNs memory of previous inputs/outputs, {which is useful for sequential processing and helps} in exploiting local and global correlations. This makes them ideal for applications with time-series data  \cite{8663966}.
\item Long Short-Term Memory (LSTM) Networks: {Although RNNs have memory, their memory is short}. {As a special type of RNNs,} LSTM networks {feature gated} memory cells which {extends their} memory to longer sequences. Due to this, LSTM networks have been used for channel estimation in channels with memory \cite{8663966}. 
\end{itemize}




{\subsection{Limitations in NG Networks}
{The aforementioned NNs can be employed to build encoding and decoding layers (an autoencoder), that produce a different representation of data and reconstruct it from the new representation, respectively. This can then be used to realize physical layer processing tasks such as signal design, channel estimation, CSI feedback, modulation, and coding} \cite{8663966}.

{However, classical DNNs encounter limitations in fully meeting the demands of NG networks. MLPs have limited ability to extract deep features from raw data, and their performance on sequential data is poor. This leads to challenges in generalizing and transferring learned knowledge between different scenarios, hindering their ability to seamlessly adapt and perform in a dynamically changing environment which is common in wireless networks. While CNNs, RNNs, and LSTM networks are more adaptable spatially/temporally to local/short-term changes, they have limited capability to effectively exploit global/long-term dependencies in sequential data, which is crucial for capturing the intricate patterns and dynamics inherent in NG wireless networks. In addition to this, training recurrent networks (RNN and LSTM) suffers from challenges related to convergence, vanishing gradients, and parallelization, which limits their usefulness in latency-sensitive applications.}

{These limitations underscore the need for innovative enhancements, alternative architectures, and hybrid approaches to effectively address the evolving requirements of NG networks.  
Recently, attention-based DL realized using transformer networks was proposed and shown to achieve remarkable performance gains in various CV and NLP applications compared to classical DL }\cite{10.5555/3295222.3295349}. Next, we present the basic architecture of transformers and discuss {their potential compared to} classical DNNs. The potential of the TMAE, a {transformer-based} autoencoder, for NG networks, is discussed afterward.

\section{Potential of Transformer-Based NNs}
A transformer is an NN architecture proposed originally for NLP \cite{10.5555/3295222.3295349}. Owing to their remarkable ability to capture complex patterns and relationships in data, transformers have been adapted for various applications, including CV and wireless communications. This is because they have several advantages over classical DNNs. First, transformers employ the attention mechanism, which allows them to dynamically weigh the importance of {segments in the data, which allows them to capture short- and long-term dependencies in data}. This is particularly beneficial in time series analysis where extended sequences of data points are involved, {enhancing their capability of generalizing and adapting to changing environments.} Second, transformers take advantage of parallel processing, which significantly improves their efficiency. Finally, pre-trained transformer models, such as BERT \cite{10.5555/3295222.3295349}, can be fine-tuned for customized objectives using a small dataset {(realizing transfer learning)}. These advantages make transformers attractive for challenging DL tasks, including wireless communication applications. This section delves into the workings of transformers and examines their key components. It also examines the architecture of the TMAE.

\subsection{Transformer Architecture}
 There is a variety of transformer architectures depending on the application. However, all architectures share the same fundamental principle: the attention mechanism. The main components of transformers, as shown in Fig. \ref{fig:Transformer Architecture.}, are input embedding, positional encoding, and multi-head attention, which are discussed next.

{\bf Input Embedding:} First, the input vector is segmented and projected into the embedding space (usually of higher dimension than the input). This can be achieved using a single convolutional layer for example. The result of this step is a representation in the embedding space of segments of the input vector, each of which is characterized by a position. 


\begin{figure} 
  \centering
  \includegraphics[scale = 0.5]{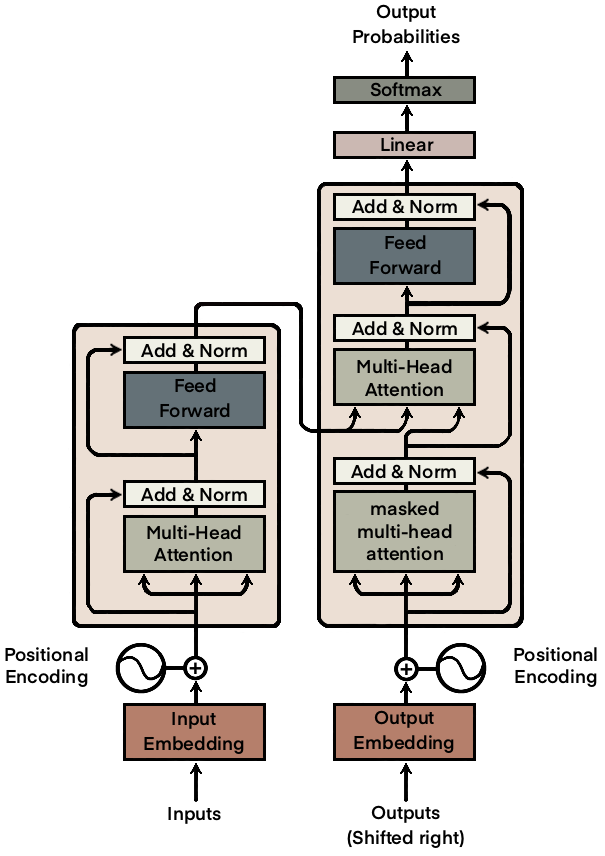}
  \caption{ Transformer Architecture (reproduced from \cite{10.5555/3295222.3295349}).}
  \label{fig:Transformer Architecture.}
\end{figure}  

{\bf Positional Encoding:} Since the transformer encoder has no recurrence (unlike RNN), it is essential to add segment position into the data. This is accomplished with positional encoding. There are numerous methods for positional encoding, one of which employs trigonometric functions. In this case, each odd indexed segment is encoded using samples of a cosine function with a frequency that depends on the position, effectively encoding this positional information in the generated vector. Similarly, samples of a sine function are used to encode the positions of even indexed segments. These positional encoding vectors are then added to the input embeddings of their respective segments.

{\bf Multi-head Attention:} Multi-head attention is the most important component of transformers and plays a crucial role in quantifying the relationships between the inputs. This is achieved using the self-attention mechanism, which relates the inputs with different positions in a sequence to provide a comprehensive representation of that sequence. Let the resulting vector obtained after position encoding of segment $i\in\{1,\ldots,n\}$ be denoted by a vector $\mathbf{x}_i$, and $\mathbf{X}=[\mathbf{x}_1,\ldots,\mathbf{x}_n]$. As shown in Fig. \ref{fig:Atten}, three NNs are used to generate three matrices from $\mathbf{X}$ which are the Key $\mathbf{K}$, the Query $\mathbf{Q}$, and the Value $\mathbf{V}$, each column of which represents one segment. Then, an attention map is generated by calculating the product $\mathbf{Q}^T\mathbf{K}$, and applying a soft-max function to map the resulting values to probabilities. The probabilities are then used as weights multiplied by $\mathbf{V}$ to obtain a self-attended feature map for $\mathbf{X}$. 
To construct multi-head attention, this attention mechanism is applied multiple times in parallel, and the resulting outputs are concatenated and projected again to obtain the final result. The rationale behind using multiple attention blocks is to enable the attention function to extract information from different perspectives (queries) and capture the complex relationships between segments. Residual connections and layer normalization are used to improve stability and performance, and a feed-forward NN is used to extract more complex features. 

To attend to relations between previous transformer outputs and current inputs, a similar mechanism is applied to generate self-attended feature maps from previous outputs, and then to generate an attention map between the current input and previous outputs {(right-most blocks in Fig. }\ref{fig:Transformer Architecture.}). Finally, the result is converted to a representation that depends on the task at hand (translation, prediction, classification, etc.).

\begin{figure} 
  \centering
  \includegraphics[scale = 0.4]{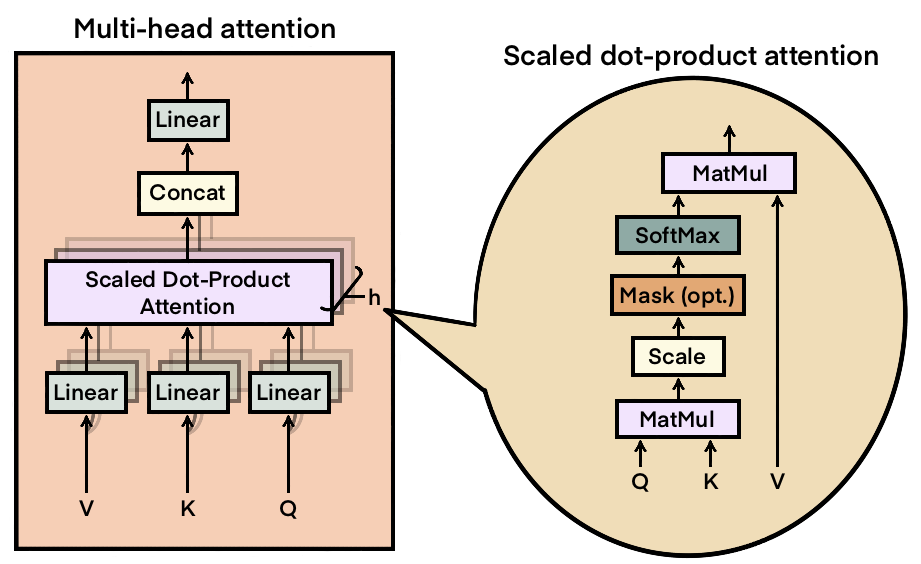}
  \caption{A multi-head attention block {where the scaled dot-product attention is applied $h$ times.}}
  \label{fig:Atten}
\end{figure}  

Based on this architecture, the authors of \cite{ViT} proposed the so-called Vision Transformer (ViT) for CV applications. Several other transformer-based architectures were proposed including the TMAE \cite{9879206} discussed next.

\begin{figure} 
  \centering
  \includegraphics[scale = 0.45]{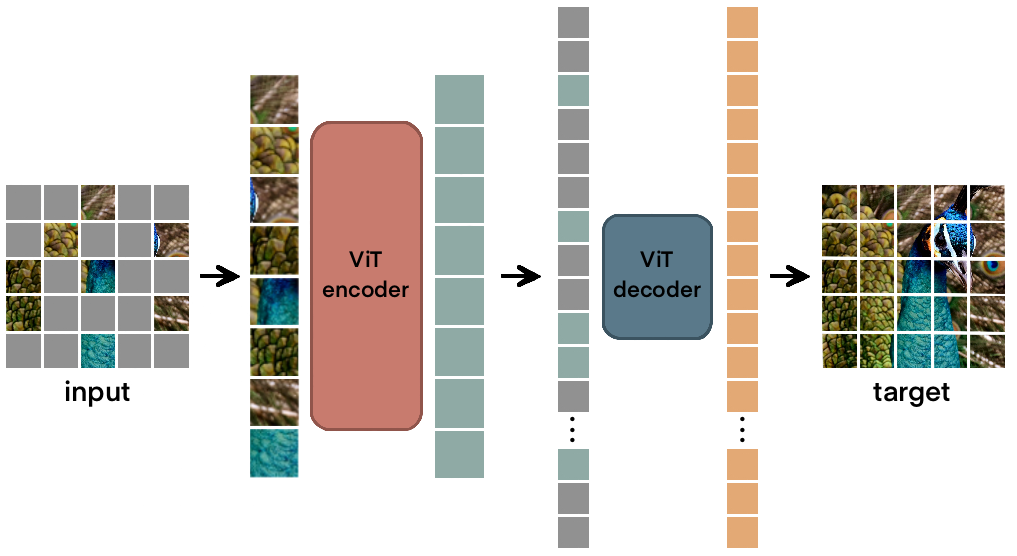}
  \caption{Masked Auto-encoder Architecture: Encoding is performed on the small subset of visible patches. The masked portions of the image are added after the encoder, and a decoder reconstructs the original image from the complete set of encoded patches and mask tokens \cite{9879206}.}
  \label{fig:MEA}
\end{figure}

\subsection{The TMAE Architecture}
A TMAE is a transformer-based architecture that aims to reconstruct data using only partial observations. Fig. \ref{fig:MEA} {shows an example where the data is an image and the transformer-based architecture is a ViT}. 
The TMAE consists of an encoder and a decoder.

{\bf TMAE encoder:} The encoder in a {TMAE} is designed to convert {partial observations} of the input data into a latent representation. It is realized using a transformer-based NN applied on data with masked segments. The transformer-based NN embeds the observed segments, adds their positional information, and then passes the result through multiple multi-head attention and DNN blocks as explained above (Fig. \ref{fig:Transformer Architecture.}). 

%

{\bf TMAE decoder:} The TMAE decoder uses the latent representation and the positions of the masked segments to reconstruct the original data as shown in Fig. \ref{fig:MEA}.

Note that the TMAE encoder is usually narrower than the TMAE decoder since the former operates on the observed segments only whereas the latter operates on both the observed and masked ones. Also, the TMAE encoder is usually deeper than the TMAE decoder because it needs to learn the semantics and correlations of the data from the observed segments. Hence, the encoder and decoder are asymmetric. 

{The potential of TMAE in NG networks is highly promising, offering a novel paradigm to address complex challenges. By amalgamating the power of transformers with the reconstruction capabilities of an autoencoder, the TMAE framework can efficiently process sequential data and capture intricate temporal dependencies inherent in wireless channels. This makes it applicable for enhancing resource efficiency, adaptability, and robustness in tasks such as signal processing, channel estimation, and resource allocation, contributing to the evolution of resilient and high-performance wireless networks.}

\subsection{Transformer Challenges}
{The superiority of transformer-based NNs comes at the cost of some challenges. Their parallel nature increases the resources needed for them to run. Transformers also require a substantial amount of labeled data for training, often necessitating pre-training on large corpora before fine-tuning on the target task. Additionally, fine-tuning pre-trained transformers on specific tasks requires careful balancing to avoid catastrophic forgetting and retain previously learned knowledge }\cite{10210052}.{ Note however that some of these challenges are shared with classical DNNs. A comparison between classical DNNs and transformer-based NNs is given in Table }\ref{table:dlr1_comparison},{ providing a concise overview of their strengths and limitations, which helps in understanding their suitability for different tasks.}

	 \begin{table*}[t]
	 	\centering
	 	\caption{{Comparative Evaluation of Deep Learning Techniques Based on Key Characteristics}}
	 	\label{table:dlr1_comparison}
	 	\begin{tabular}{|l|c|c|c|c|c|}
	 		\hline
	 		\textbf{Aspect} & \textbf{MLP} & \textbf{CNN} & \textbf{RNN} & \textbf{LSTM} & \textbf{Transformer} \\
	 		\hline
	 		Architecture & Fully Connected & Convolutional & Recurrent & Recurrent & Attention-based \\
	 		\hline
	 		Sequence Modeling & No & Partial & Yes & Yes & Yes \\
	 		\hline
	 		Long-term Dependencies & Limited & Limited & Yes & Moderate & Improved \\
	 		\hline
	 		Parallel Processing & No & Yes & No & No & Yes \\
	 		\hline
	 		Spatial Invariance & No & Yes & No & No & No \\
	 		\hline
	 		Parameter Sharing & No & Yes & Yes & Yes & Yes \\
	 		\hline
	 		Interpretability & Low & Moderate & Low & Moderate & Moderate \\
	 		
	 		\hline
	 		
	 		Computational Complexity & Low & Moderate & Low & High & High \\
	 		\hline
	 		Memory and Storage Requirements & Low & Low & Moderate & High & High \\
	 		\hline

	 	\end{tabular}
	 \end{table*}

In the next section, we provide a case study illustrating how a TMAE enhances the compression rate of existing compression schemes.

\section{Case Study: TMAE-Enhanced Compression}
\label{Sec:Case_Study}

{In some communication applications, communicating nodes have limited computational resources and are deployed in resource-constrained environments. Examples include Unmanned Aerial Vehicle (UAV) and Internet of Things (IoT) applications. Yet, the demand for transmitting vast amounts of data (whether it be images, sensor readings, or text) persists. Given the limited resources in these systems, transmitting vast amounts of data becomes challenging. Fortunately, such data often exhibits inherent correlations between its segments. For instance, in an image, certain patches can be inferred from their neighboring patches due to the visual structure and semantics of the scene. Similarly, in a sequence of sensor readings, some values can be predicted based on the preceding and succeeding values. Thus, one can omit certain segments of the data, leveraging the correlations to infer these segments at the receiver's end. This approach not only reduces the amount of data to be transmitted but also requires less processing resources at the transmitter, making it particularly advantageous for systems like UAVs capturing and transmitting real-time images or IoT devices sending frequent sensor updates.}

\subsection{Example: UAV with TMAE-Enhanced Compression}

\begin{figure*}

  \centering
  \includegraphics[width=.9\linewidth]{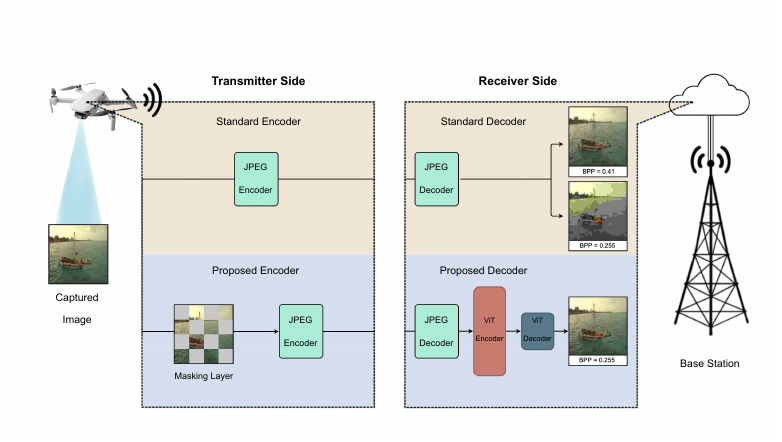}
  \caption{{Remote UAV imaging: A qualitative comparison between the conventional JPEG compression and the proposed JPEG-TMAE compression scheme.}}
  \label{fig: Model_arch}
\end{figure*}

{Consider a UAV operating in a distant, hard-to-reach region, tasked with gathering essential information about the landscape through imaging. The primary challenge for this UAV is not merely capturing the images but efficiently transmitting them back to a base station (BS). In this scenario, the communication channel quality is poor, limiting the amount of information that can be sent. This requires compressing the image aggressively, which can compromise its meaningful interpretation. Conventional compression techniques may fail in this scenario. Fig.} \ref{fig: Model_arch} {shows an example where a standard compression scheme produces a very low-quality image when the compression (in bits per pixel (BPP)) is low.}

{TMAE-enhanced compression offers a distinct advantage. Once an image is captured, the UAV segments it into non-overlapping patches and randomly masks some of them, relying on the ability of the BS to infer them from unmasked patches due to the potentially high correlation between patches. Once the patches are masked and removed, the remaining patches are stacked to form a condensed image, which is then further compressed using a standard compression algorithm. Upon receipt at the BS, the compressed image is decompressed to retrieve the stacked patches, and the TMAE, trained to handle random masks, reconstructs the masked patches to reproduce the complete image. For accurate reconstruction, the TMAE requires information about the location of the masked patches. This mask can be obtained by sharing the seed of the random number generator used to generate the mask (or can be stored at the receiver if it is deterministic). While the resulting image might not be of the highest quality, it conveys essential and interpretable information from the UAV despite the communication challenges, as can be seen in Fig.}~\ref{fig: Model_arch}.

{Next, we examine the proposed scheme quantitatively when JPEG is used as the standard compression algorithm.}

\subsection{JPEG-TMAE Compression}

{The proposed JPEG-TMAE compression scheme is a combination of the widely-used JPEG compression and the TMAE, as shown in Fig. }\ref{fig: Model_arch}.{ To implement this, we use patches of size $16\times 16$ pixels (as an example). Then, we use a masking ratio of $R_{\rm mask}=0.67$ (i.e., $\nicefrac{2}{3}$ of the image is masked) and random masking with a seed that is shared between the UAV and the BS. Thus, the compression gain achieved by the TMAE is the ratio between the size of the original and the stacked patches $(1-R_{\rm mask})$. The remaining patches are stacked and compressed using JPEG. As such, the overall compression rate equals the masking ratio multiplied by the JPEG compression rate. Upon receiving the compressed image, the BS decompresses it using JPEG, and then uses a pre-trained TMAE to reconstruct the image. }

{Leveraging the pre-trained transformers-based model from Facebook AI Research (FAIR)} \cite{9879206},{ we bypassed extensive training, focusing instead on architectural modifications tailored to our compression goals. This strategic approach ensured an optimal balance between computational efficiency and performance in the described UAV communication scenario.}

{Fig. }\ref{fig: Model_arch}{ shows a comparison where JPEG and the proposed JPEG-TMAE schemes are used to compress an image, using a compression rate of $0.41$ and $0.255$ BPP for JPEG, and a compression rate of $0.255$ BPP for the JPEG-TMAE scheme. The JPEG scheme fails at the compression rate $0.255$ BPP, contrary to the proposed JPEG-TMAE which reconstructs the image with a much better quality at the same overall compression rate (a quality comparable with JPEG at $0.41$ BPP).}


{For a quantitative comparison, we can use the structural similarity measure (SSIM) to assess the quality of the reconstructed images }\cite{SSIM}.{ SSIM gauges the similarity between the input and output images, factoring in the interpixel dependencies of closely situated pixels. Recognized as a full reference metric, SSIM measures image quality using an uncompressed or noise-free initial input image as its reference. Thus, we examine the SSIM of the proposed JPEG-TMAE on the Kodak dataset} \cite{kodak-dataset}{ (often used to compare compression methods).  Through a methodical process of iterating the masking ratio and adjusting the JPEG compression rate for every iteration, we were able to delineate an optimized performance spectrum across all examined masking ratios. We compare with several leading models, specifically mbt2018 (CNN-Based)} \cite{minnenbt18}, {cheng2020-anchor (CNN-Based) }\cite{cheng2020image},{ and ConvLSTM (CNN-LSTM Based) }\cite{ConvLSTM}.

{Fig} \ref{fig:results} {displays the SSIM in relation to the overall compression rate. Notably, the JPEG-TMAE not only outperforms JPEG, particularly at lower compression rates, but it also exhibits superior performance compared to leading models such as mbt2018, cheng2020-anchor, and ConvLSTM at these rates. In the realm of moderate compression rates, it matches the performance of these models. An added advantage is its design approach: while the aforementioned methods are based on autoencoders and thus require part of the DNN architecture to be incorporated at the UAV}---{increasing its complexity, our JPEG-TMAE seamlessly integrates the entire NN onto the BS, thus simplifying processing at the UAV. Note that while the ConvLSTM method performs best at higher compression rates, we expect a similar transformer}--{based architecture to perform even better.}

{In summary, this approach is highly suitable for communication applications with limited resources at the transmitter side (UAVs, IoT, etc.). Note that the results in Fig.} \ref{fig:results} {can be further improved using dynamic masking to ensure sufficient correlation between masked/unmasked patches, leading to either a lower compression rate or better image quality.}

\begin{figure}
\centering
\tikzset{every picture/.style={scale=.8}, every node/.style={scale=.9}}
%
%
\definecolor{mycolor1}{rgb}{0.00000,0.44700,0.74100}%
\definecolor{mycolor2}{rgb}{0.85000,0.32500,0.09800}%
\definecolor{mycolor3}{rgb}{0.92900,0.69400,0.12500}%
\definecolor{mycolor4}{rgb}{0.49400,0.18400,0.55600}%
\definecolor{mycolor5}{rgb}{0.46600,0.67400,0.18800}%
\definecolor{mycolor6}{rgb}{0.30100,0.74500,0.93300}%
\definecolor{mycolor7}{rgb}{0.63500,0.07800,0.18400}%
\begin{tikzpicture}

\begin{axis}[%
width=4.5in,
height=3.5in,
scale only axis,
xmin=0.1,
xmax=0.5,
xlabel style={font=\color{white!15!black}, at={(axis cs: 0.3,0.1)}},
xlabel={Bit per pixel (BPP)},
ymode=log,
ymin=0.1,
ymax=0.5,
yminorticks=true,
ytick = {0.1,0.15,0.2, 0.25, 0.3, 0.35, 0.4, .45, .5},
yticklabels = {0.1,0.15,0.2, 0.25, 0.3, 0.35, 0.4, 0.45, 0.5},
ylabel style={font=\color{white!15!black}, at={(axis cs: 0.1,0.25)}},
ylabel={SSIM},
axis background/.style={fill=white},
xmajorgrids,
ymajorgrids,
yminorgrids,
legend style={at={(axis cs: 0.5,0.1)}, anchor=south east, legend cell align=left, align=left, draw=white!15!black}
]
\addplot [color=red, line width=1.0pt]
  table[row sep=crcr]{%
0.26616974914966	0.216819795287559\\
0.266302614795918	0.217076911094681\\
0.278984640731293	0.249056392831938\\
0.303139615221088	0.301838144106932\\
0.328636532738095	0.347287831829525\\
0.566931069302721	0.557600654247583\\
};
\addlegendentry{JPEG}

\addplot [color=black, line width=1.0pt]
  table[row sep=crcr]{%
0.131384194302721	0.0957299056703332\\
0.143541400935374	0.149210727406744\\
0.158814306972789	0.179399724771426\\
0.17784731079932	0.208815158187379\\
0.188436702806122	0.230824914521954\\
0.219334608843537	0.2776170957478\\
0.258576477465986	0.324522948655854\\
0.296463116496599	0.365897396562047\\
0.34402237457483	0.409028052413993\\
0.442017431972789	0.464381009398488\\
0.484634088010204	0.480418601139433\\
0.561330782312925	0.503685930965175\\
};
\addlegendentry{JPEG-TMAE (Optimized)}

\addplot [color=green, line width=1.0pt]
  table[row sep=crcr]{%
0.15583462185329858    0.10005479852202352\\
0.1617889404296875     0.13439492852385929\\
0.16532474093967017    0.15059053639389894\\
0.16795010036892366    0.15883068032018086\\
0.16986423068576387    0.16415489353010065\\
0.17078823513454863    0.16656789836941904\\
0.17131720648871526    0.16811935272638226\\
0.17161220974392363    0.16865256504237\\
0.17186652289496526    0.16927173673077942\\
0.17205047607421875    0.16978527558660705\\
0.1721369425455729     0.17018323184257259\\
0.17225731743706596    0.1701982345665436\\
0.17243872748480904    0.17030083823028472\\
0.17266591389973954    0.17049914467515284\\
0.2577395968967014     0.26311950422145176\\
0.3071738349066841     0.32328938829565285\\
0.33673434787326384    0.3824002583473112\\
0.3608957926432292     0.4207165501303598\\
0.3771396213107639     0.4428826127495627\\
0.38805389404296875    0.45852518054993957\\
0.3957095675998265     0.46985682074589286\\
0.4020572238498264     0.47701610728861965\\
0.40517340766059035    0.4811334995592545\\
0.4079615275065105     0.4845167843088319\\
0.40952131483289916    0.48739381847968577\\
0.4108784993489584     0.4903266787751413\\
0.412324693467882      0.4924518318067699\\
0.4134462144639757     0.49421304458665083\\
};
\addlegendentry{ConvLSTM (CNN-LSTM Based)\cite{ConvLSTM}}

\addplot [color=brown, line width=1.0pt]
  table[row sep=crcr]{%
0.17167522 0.16625049\\
0.21742701 0.27015172\\
0.27555805 0.34790647\\
0.31232452 0.38280889\\
0.39392401 0.43696703\\
0.46371333 0.46766172\\
0.58388265 0.50097406\\
0.77102294 0.53066724\\
1.01181511 0.54945989\\
};
\addlegendentry{cheng2020-anchor(CNN based)\cite{cheng2020image}}

\addplot [color=blue, line width=1.0pt]
  table[row sep=crcr]{%
0.17003716 0.16144037\\
0.21402232 0.25665099\\
0.26508416 0.31950167\\
0.29616631 0.34490801\\
0.36334144 0.38455074\\
0.45891232 0.46451866\\
0.57301500 0.49298646\\
0.74923282 0.52114487\\
0.86121707 0.53466851\\
};
\addlegendentry{mbt2018(CNN Based)\cite{minnenbt18}}

\end{axis}

\end{tikzpicture}%
\caption{{Comparison of the JPEG-TMAE compression scheme with conventional JPEG compression and state-of-the-art models on the Kodak dataset.}}
\label{fig:results}
\end{figure}
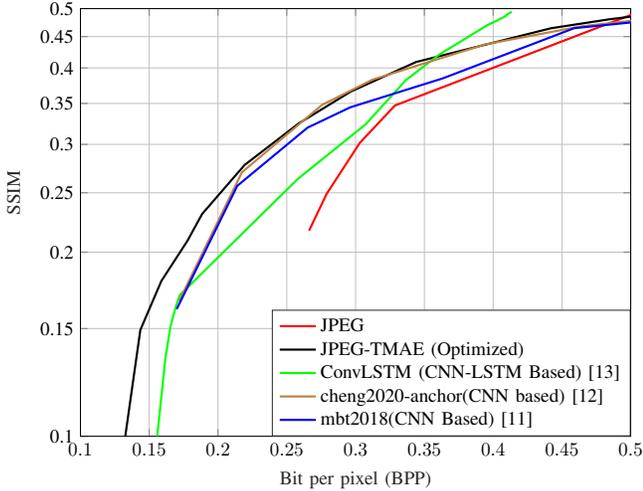

\section{Applications in NG Networks}
As demonstrated above, the TMAE can be used efficiently to reduce the size of an image and reconstruct it using its semantics. This improves source coding and consequently increases the throughput of wireless communications. In this section, we discuss applications of the TMAE in various areas of wireless communication. 

\subsection{Semantic Source and Channel Coding}
In Sec. \ref{Sec:Case_Study}, we present an application of the TMAE for image compression. {In general, the source coding benefits of TMAEs extend beyond image compression.} The proposed TMAE-enhanced compression can be extended to video data for instance. However, for video data, the masks will have both space and time dimensions. Similarly, TMAEs can be combined with standard compression schemes as in Sec. \ref{Sec:Case_Study} to improve the compression of text or audio data. {This source coding potential can increase the throughput of NG networks.}

{In addition to source coding, TMAEs can be beneficial in channel coding. Autoencoders in general have been used in the literature for various transmission processing applications including channel coding and modulation }\cite{sun2019application}.{ As an autoencoder with attention, the TMAE can be superior in such tasks. In particular, the attention mechanism gives TMAEs an edge in joint source-channel coding applications }\cite{jiang2020joint}.{ Also, their ability to learn long-term dependencies can be beneficial for communications over changing communication environments. Combining TMAEs with various NN architectures or classical coding schemes (LDPC, polar codes) to achieve superior performance is a promising research direction.}

  


\subsection{Channel Estimation and Prediction}
One stringent requirement in NG networks is to guarantee extremely low latency. One way to achieve this is to shrink the packet size to reduce the decoding transmission and computation time. However, the pilot sequences used for channel estimation increase the packet size. In addition, in dynamic environments where users are highly mobile, the channels change rapidly which requires more frequent pilot signal transmission. {Researchers used NNs for channel estimation/prediction to mitigate this problem. However, the complexity and dynamicity of wireless channels limit the capability of existing NNs to improve channel estimation performance. 
We envision the TMAE to be a strong candidate for reducing pilot signal transmission overhead due to its superiority in reconstructing signals from partial observations.}

For example, in OFDM, the channel time-frequency response can be envisioned as a 2D image (with colors representing the real and imaginary parts) \cite{soltani2019deep}. {The TMAE can be exploited for reducing the number of estimated channels, by estimating the channels of some time-frequency resource blocks, and inferring the rest using the TMAE by treating them as masked patches in the 2D image.} This idea can also be applied to multiple antenna systems where the channels' spatial, temporal, and spectral characteristics need to be estimated. 
Another example {is channel estimation for} reconfigurable intelligent surfaces (RIS), which are surfaces that can be mounted on buildings to reflect signals in a desired direction and improve channel quality. {Despite their advantages, they require} large channel estimation overheads. 
A TMAE can be used to reduce this overhead, {by estimating parts of the RIS channel matrices, and exploiting the spatial correlation in these matrices to reconstruct the rest of the channel using a TMAE.}

\subsection{Privacy and Security}
Data collection is a major concern for NG networks. For applications involving DL, collaborative learning in the cloud provides a means for processing large amounts of data collected from distributed nodes. However, this comes at the expense of privacy when the application involves using sensitive data or when the trained model is shared. Intelligent {NG networks should adopt} privacy-by-design approaches that are service-oriented and privacy-preserving. Several privacy challenges can be addressed using a TMAE to build generative models to construct privacy-preserving datasets, where data privacy and utility can be ensured simultaneously. {For instance,} a TMAE can be used to encrypt user data before it is transmitted over the network, {so that it is only accessible to authorized users who have access to} the appropriate mask. 

{A TMAE can also be useful for detecting} potential for attacks on the network infrastructure itself. {It can be used to analyze network traffic and detect anomalies or suspicious patterns with higher accuracy than classical DNNs.} In general, by leveraging the power of the TMAE, more effective and efficient solutions for securing networks and protecting users' data can be developed.

\section{Challenges and Opportunities of TMAE in NG Networks}

{Although the integration of TMAE into NG networks holds transformative promise, it also brings to the forefront a series of multifaceted challenges and open problems. These intricacies arise from the combined complexity of transformer models and dynamic wireless networks. In this section, we discuss the current challenges and future research directions of TMAE in NG networks.}

\begin{itemize}
\item {\bf Computational resources:} { Although the TMAE benefits from parallel processing to achieve remarkable performance, this comes at the cost of computational resources (Graphics Processing Units (GPUs)). In the case study above, we benefited from the ability to migrate the TMAE processing to the BS which can be equipped with a GPU. However, if the processing has to take place at a resource-constrained device for latency considerations (such as an autonomous vehicle), this will require this device to be equipped with expensive hardware (computational resources and memory) which may not always be practical. The lack of such hardware will reduce the performance or increase the latency of TMAEs, which limits their use in real-time applications. Resolving this limitation requires innovative techniques for distributing the TMAE between devices and edge clouds (federated transformer-based NN). Moreover, like other DNNs, the scalability of the TMAE for use in large-scale communication networks presents challenges in terms of computational resources. }

\item {\bf Energy consumption:} {In mobile and battery-operated wireless devices, energy efficiency is a critical concern. The computational complexity and memory requirements of TMAEs (and more generally transformer-based NNs) can lead to high energy consumption, making them less energy-efficient for resource-constrained devices. For energy-critical wireless communication networks, researchers need to explore model compression techniques, data quantization methods, and hardware acceleration to reduce energy consumption while maintaining satisfactory performance.}

\item {\bf Pre-trained models:} {Pre-trained transformers have shown extraordinary performance in CV and NLP, and excellent adaptability between different applications therein. It is expected that this property of transformers transfers well to wireless communications applications. However, existing pre-trained transformer models (trained on images or text) are not optimized for wireless communications data, and their fine-tuning on such data may not provide the best performance. On the other hand, transformer models that are trained on wireless communications data do not exist nowadays. There is a great opportunity for researchers to develop pre-trained transformer models that are trained on wireless communications data (such as channels and resource allocation data), and to demonstrate the adaptability of these models within different applications in wireless communications with minimal training requirements. This also highlights another opportunity, which is the development of a large dataset of wireless communications data for such purposes. The availability of diverse and abundant data from wireless communications applications will enable efficient training and improved generalization within wireless communications applications.}
\end{itemize}



{In summary, incorporating TMAE into future wireless communication systems holds unparalleled potential, but addressing these intertwined challenges and open problems is fundamental to their seamless deployment. By collectively tackling these intricacies, the wireless communication community can usher in a new era of adaptive, efficient, and secure communication networks.}

\section{Conclusion}
In this article, we discussed the limitations of using classical deep-learning methods in wireless networks. We then presented the architecture of transformers and masked autoencoders and discussed their distinct capabilities compared to traditional deep-learning methods. We also showed an application of transformer-masked autoencoders in data compression, which yielded a significant improvement compared to classical approaches. We explored and presented some applications and open research problems, where transformer-masked autoencoder-based solutions can be developed to produce intelligent wireless communication systems. In general, it is expected that transformer-based neural networks will play an important role in next-generation communication networks, and there are many challenges and opportunities in this area for the research community to explore.

\bibliographystyle{IEEEtran}
\bibliography{ref}

\begin{IEEEbiography}
   [{\includegraphics[width=1in,height=1.25in, clip, keepaspectratio]{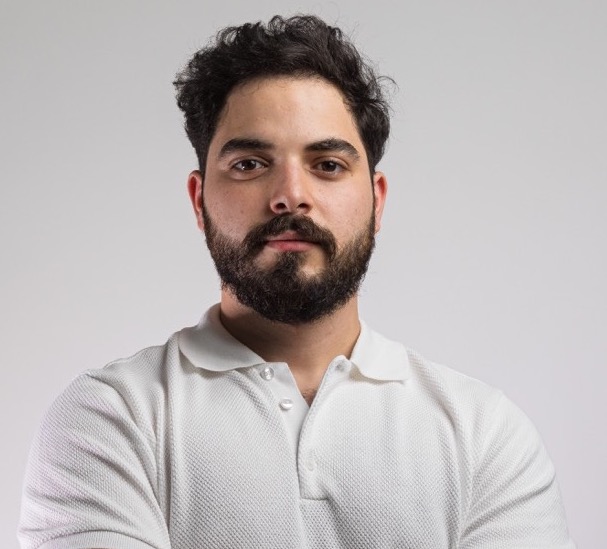}}]  {Abdullah Zayat} ({S'21}) received B.S. degree in electrical engineering from King Fahd University of Petroleum and Minerals (KFUPM), Saudi Arabia, in 2021. Between 2019 and 2020, he was a Research Intern at King Abdullah University of Science and Technology (KAUST). He then obtained his M.A.Sc. degree in electrical engineering from the University of British Columbia (UBC), Canada, in 2023. Currently, he is a PhD student in Electrical Engineering at UBC. His research interests focus on signal processing, communications, and the application of machine/deep learning in communications.
\end{IEEEbiography}

\begin{IEEEbiography}[{\includegraphics[width=1in,height		=1.25in,clip,keepaspectratio]{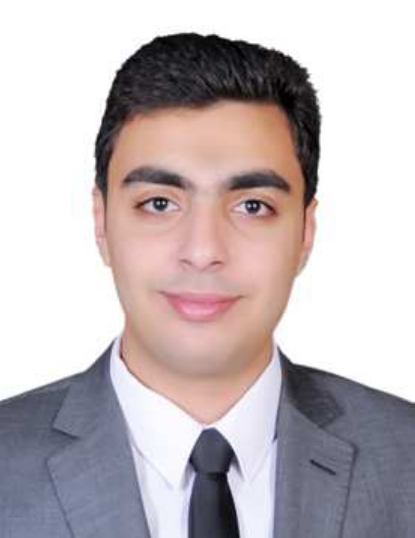}}] {Mahmoud A. Hasabelnaby} ({S'20})  received the B.S. (Hons.) and M.S. degrees in electronics and electrical communications engineering from Menoufia University, Menouf, Egypt, in 2014 and 2019, respectively. He is currently working toward a Ph.D. degree in electrical engineering at the University of British Columbia, Okanagan Campus, Kelowna, Canada. He is on leave from the Faculty of Electronic Engineering, Menoufia University, Menouf,  Egypt. From 2016 to 2018, he was a Research Assistant at the National Telecommunications Regularity Authority, Egypt. His research interests include wireless communications, information theory, ML/AI, end-to-end cloud-native software development, and next-generation wireless access networks. 
\end{IEEEbiography}

\begin{IEEEbiography}
[{\includegraphics[width=1in,height=1.1in,clip,keepaspectratio]{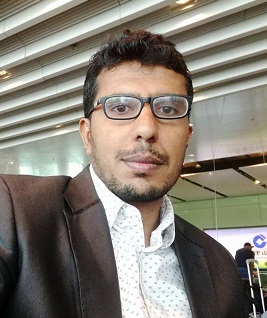}}]{Mohanad Obeed} received the B.Eng. degree in computer and communication engineering from Taiz University, Taiz, Yemen, in 2008, the M.Sc.  and the Ph.D. degree in electrical engineering from King Fahd University of Petroleum and Minerals (KFUPM), Dhahran,  Saudi Arabia, in 2016 and 2019, respectively. From July 2017 to July 2019, he was a visiting researcher at King Abdullah University of Science and Technology (KAUST) under the supervision of Mohamed-Slim Alouini.  He was a Postdoctoral Research Fellow with the School of Engineering at the University of British Columbia, Canada, from 2019 to 2022. He joined Carleton University, as a postdoctoral research fellow, in 2023. His research interests include satellite communication, 5G and 6G networks, channel estimation, deep learning, and federated learning. \end{IEEEbiography} 

\begin{IEEEbiography}[{\includegraphics[width=1in,height=1.30in,clip,keepaspectratio]{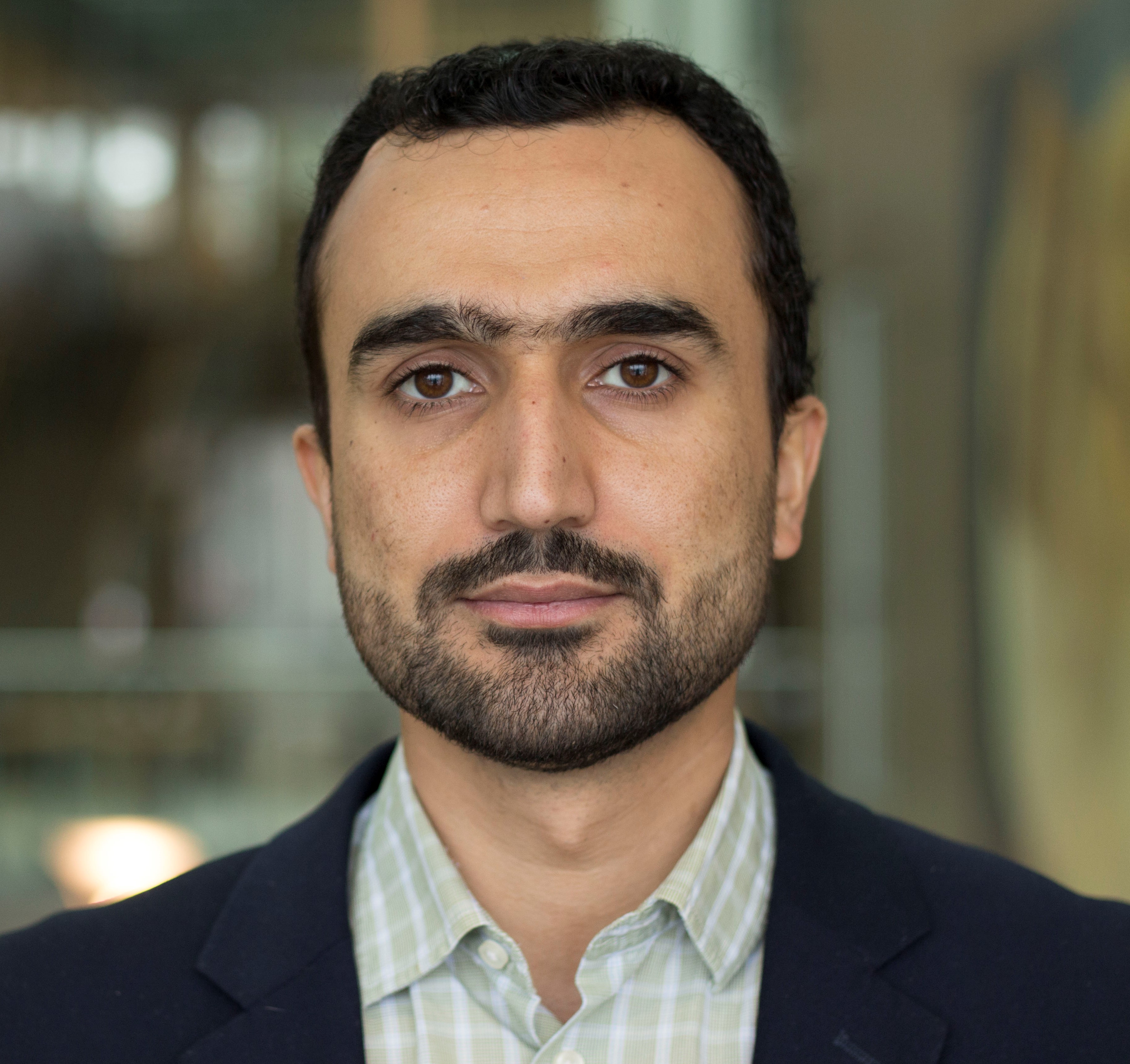}}]{Anas Chaaban} (S'09 - M'14 - SM'17) received the Ma{\^i}trise {\`e}s Sciences degree in electronics from Lebanese University, Lebanon, in 2006, the M.Sc. degree in communications technology and the Dr. Ing. (Ph.D.) degree in electrical engineering and information technology from the University of Ulm and the Ruhr-University of Bochum, Germany, in 2009 and 2013, respectively. From 2008 to 2009, he was with the Daimler AG Research Group On Machine Vision, Ulm, Germany. He was a Research Assistant with the Emmy-Noether Research Group on Wireless Networks, University of Ulm, Germany, from 2009 to 2011, which relocated to the Ruhr-University of Bochum in 2011.  He was a PostDoctoral Researcher with the Ruhr-University of Bochum from 2013 to 2014,  and with King Abdullah University of Science and Technology from 2015 to 2017. He joined the School of Engineering at the University of British Columbia as an Assistant Professor in 2018. His research interests are in the areas of information theory and wireless communications. \end{IEEEbiography}


\end{document}